\documentclass[
    superscriptaddress,
    reprint,
    showpacs,
    amsmath,
    amssymb,
    aps,
    prb,
    floatfix
]{revtex4-1}
\usepackage{graphicx}
\usepackage{dcolumn}
\usepackage{bm}
\usepackage{braket}
\usepackage{subfigure}
\usepackage{hyperref}
\hypersetup{colorlinks=true,linkcolor=blue,citecolor=blue,urlcolor=blue}
\begin{document}
\title{
Zone-center phonons of bulk, few-layer, and monolayer 1T-TaS$_2$: Detection of the commensurate charge density wave phase through Raman scattering
}
\author{Oliver R. Albertini}
\email{ora@georgetown.edu}
\affiliation{
    Department of Physics,
    Georgetown University,
    Washington, DC 20057, USA
}
\author{Rui Zhao}
\affiliation{
    Materials Science and Engineering Department,
	The Pennsylvania State University, University Park,
	Pennsylvania 16802, USA
}
\affiliation{
	Center for 2-Dimensional and Layered Materials,
	The Pennsylvania State University, University Park,
	Pennsylvania 16802, USA
}

\author{Rebecca L. McCann}
\affiliation{
    Department of Physics,
    Georgetown University,
    Washington, DC 20057, USA
}
\author{Simin Feng}
\affiliation{
	Center for 2-Dimensional and Layered Materials,
	The Pennsylvania State University, University Park,
	Pennsylvania 16802, USA
}
\affiliation{
    Department of Physics,
	The Pennsylvania State University, University Park,
	Pennsylvania 16802, USA
}
\author{Mauricio Terrones}
\affiliation{
    Materials Science and Engineering Department,
	The Pennsylvania State University, University Park,
	Pennsylvania 16802, USA
}
\affiliation{
	Center for 2-Dimensional and Layered Materials,
	The Pennsylvania State University, University Park,
	Pennsylvania 16802, USA
}
\affiliation{
    Department of Physics,
	The Pennsylvania State University, University Park,
	Pennsylvania 16802, USA
}
\author{James K. Freericks}
\affiliation{
    Department of Physics,
    Georgetown University,
    Washington, DC 20057, USA
}
\author{Joshua A. Robinson}
\affiliation{
    Materials Science and Engineering Department,
	The Pennsylvania State University, University Park,
	Pennsylvania 16802, USA
}
\affiliation{
	Center for 2-Dimensional and Layered Materials,
	The Pennsylvania State University, University Park,
	Pennsylvania 16802, USA
}
\author{Amy Y. Liu}
\affiliation{
    Department of Physics,
    Georgetown University,
    Washington, DC 20057, USA
}
\date{\today}
\begin{abstract}
We present first-principles calculations of the vibrational properties of the transition metal dichalcogenide 1T-TaS$_2$ for various thicknesses in the high-temperature (undistorted) phase and the low-temperature commensurate charge density wave (CDW) phase.
We also present measurements of the Raman spectra for bulk, few-layer, and monolayer samples at temperatures well below that of the bulk transition to the commensurate phase.
Through our calculations, we identify the low-frequency folded-back acoustic modes as a convenient signature of the commensurate CDW wave structure in vibrational spectra.
In our measured Raman spectra, this signature is clearly evident in all of the samples, indicating that the commensurate phase remains the ground state as the material is thinned, even down to a single layer.
This is in contrast to some previous studies which suggest a suppression of the commensurate CDW transition in thin flakes.
We also use polarized Raman spectroscopy to probe c-axis orbital texture in the low-T phase, which has recently been suggested as playing a role in the metal-insulator transition that accompanies the structural transition to the commensurate CDW phase.
\end{abstract}
\maketitle

\section{\label{sec:intro}Introduction}

The 1T polymorph of bulk TaS$_2$, a layered transition-metal dichalcogenide (TMD), has a rich phase diagram that includes multiple charge density wave (CDW) phases that differ significantly in their electronic properties.
At temperatures above 550 K the material is metallic and undistorted.
Upon cooling, it adopts a series of CDW phases.
An incommensurate CDW structure (I) becomes stable at 550 K, a nearly commensurate phase (NC) appears at 350 K, and a fully commensurate CDW (C) becomes favored at about 180 K.
At each transition, the lattice structure distorts and the electrical resistivity increases abruptly.
In the C phase, 1T-TaS$_2$ is an insulator.\cite{thompson, disalvo, sipos}
Upon heating, an additional triclinic (T) nearly commensurate CDW phase occurs roughly between 220-280 K.\cite{T-phase}
While the 1T-TaS$_2$ phase diagram is well studied,\cite{T-phase, thompson, brouwer-jellinek1, disalvo, brouwer-jellinek2, wu, burk, spijkerman, sipos, yu} the exact causes of the electronic and lattice instabilities remain unclear.
Electron-electron interactions, electron-phonon coupling, interlayer interactions, and disorder may all play a role.
It has long been assumed, for example, that electron correlations open up a Mott-Hubbard gap in the C phase.\cite{fazekas}
Recent theoretical work, however, suggests that the insulating nature of the C phase may come from disorder in orbital stacking instead.\cite{millis, ritschel}

Advances in 2D material exfoliation and growth have enabled studies of how dimensionality and interlayer interactions affect the phase diagram of 1T-TaS$_2$.
Some groups have reported that samples thinner than about 10 nm lack the jump in resistivity that signals the transition to the C CDW phase.\cite{yu}
However, this may not be an intrinsic material property, as oxidation may suppress the transition.
Indeed, samples as thin as 4 nm---when environmentally protected by encapsulation between hexagonal BN layers---retain the NC to C transition.\cite{tsen}
Whether the C phase is the ground state for even thinner samples, and whether this leads to an insulating state, remain open questions.

Materials that exhibit abrupt and reversible changes in resistivity are attractive for device applications, particularly when the transition can be controlled electrically.
Ultrathin samples may exhibit easier switching than bulk materials due to reduced screening and higher electric-field penetration.
Hollander \emph{et al.} have demonstrated fast and reversible electrical switching between insulating and metallic states in 1T-TaS$_2$ flakes of thickness 10 nm or more.\cite{hollander}
Tsen and coworkers have electrically controlled the NC-C transition in samples as thin as 4 nm and identified current flow as a mechanism that drives transitions between metastable and thermodynamically stable states.\cite{tsen}
Focusing on the C phase rather than the NC-C transition, theoretical work indicates that in few-layer 1T-TaS$_2$ crystals, the existence or absence of a band gap depends on the c-axis orbital texture, suggesting an alternate approach to tuning or switching the electronic properties of this material.\cite{ritschel}

Raman spectroscopy has been a powerful tool for characterizing 2D TMDs.\cite{lee, mos2, wse2_1, wse2_2, zhao2013lattice, phonon_review, zhao-interlayer, molina-sanchez, humberto}
Because the phonon frequencies evolve with the number of layers, Raman spectroscopy offers a rapid way to determine the thickness of few-layer crystals.\cite{qiao2015substrate}
Raman spectra can also reveal information about interlayer coupling,\cite{zhao-interlayer, molina-sanchez,tan2012shear,zhang} stacking configurations,\cite{puretzky} and environmental effects like strain and doping.\cite{phonon_review}
For these materials, the interpretation of measured Raman spectra has benefited greatly from theoretical analyses utilizing symmetry principles and materials-specific first-principles calculations.
Thus far, theoretical studies of TMD vibrational properties have focused on the 2H structure favored by group-VI dichalcogenides like MoS$_2$ and WSe$_2$.\cite{phonon_review}

We present a symmetry analysis and computational study of the Raman spectrum of 1T-TaS$_2$ in the undistorted normal 1T phase as well as in the low-temperature C phase, alongside Raman measurements of the latter.
The symmetry of the 1T structure differs from that of the 2H structure, and different modes become Raman active in the normal phase.
In the C phase, a Brillouin zone reconstruction yields two groups of folded-back modes, acoustic and optical, separated by a substantial gap.
The low-frequency folded-back acoustic modes serve as a convenient signature of the C phase.
In our measured Raman spectra, this signature is evident even in a monolayer.
This is in contrast to some previous studies\cite{yu,tsen,luican-mayer,tsen2} which suggest a suppression or change in nature of the C phase transition in thin flakes.
We also investigate the effects of c-axis orbital texture on the Raman spectra and use polarized Raman measurements to narrow down the possible textures.

\section{\label{sec:structures}Description of crystal structures}
A single `layer' of TaS$_2$ consists of three atomic layers: a plane of Ta atoms sandwiched between two S planes.
In each plane, the atoms sit on a triangular lattice, and in the 1T polytype, the alignment of the two S planes results in Ta sites that are nearly octahedrally coordinated by S atoms; the structure is inversion symmetric.
The bulk material has an in-plane lattice constant of $a = 3.365$ \AA, and an out-of-plane lattice constant of $c = 5.897$ \AA.\cite{spijkerman, brouwer-jellinek1}

In the C CDW phase, two concentric rings, each consisting of six Ta atoms, contract slightly inwards towards a central Ta site ($\sim6\%$ and 3$\%$ for the first-neighbor and second-neighbor rings, respectively), forming 13-atom star-shaped clusters in the Ta planes (Fig. \ref{fig:sod-triclinic}).
The in-plane structure is described by three CDW wave vectors oriented 120$^{\circ}$ to each other, resulting in a $\sqrt{13}a\times\sqrt{13}a$ supercell that is rotated $\phi_{\mathrm{CDW}}$ = 13.9$^\circ$ from the in-plane lattice vectors of the primitive cell.
The S planes buckle outwards, particularly near the center of the clusters, inducing a slight $c$-axis swelling of the material.\cite{bovet}

The vertical alignment of these star-of-David clusters determines the so-called stacking of the C phase.
Because the electronic states near the Fermi level are well described by a single Wannier orbital centered on each cluster, the stacking arrangement of the clusters is also referred to as c-axis orbital texture.
In hexagonal stacking, stars in neighboring layers align directly atop each other.
In triclinic stacking, stars in adjacent layers are horizontally shifted by a lattice vector of the primitive cell, resulting in a 13-layer stacking sequence.
Some experiments suggest that the clusters in the C phase align in a 13-layer triclinic arrangement,\cite{wilson, williams, scruby} while others have suggested mixed\cite{nakanishi} or even disordered\cite{brouwer-jellinek1,tanda} stacking.
Interestingly, the precise stacking sequence strongly affects the in-plane transport, due to the overlap between laterally shifted Wannier orbitals in adjacent layers.\cite{ritschel, yizhi-ge}

\section{\label{sec:methods}Methods}

\subsection{\label{sec:comp-methods}Computational Methods}

Density functional theory (DFT) calculations were performed with the Quantum Espresso\cite{espresso} package.
The electron-ion interaction was described by ultrasoft pseudopotentials,\cite{vanderbilt} and the exchange-correlation interaction was treated with the local density approximation (LDA).\cite{perdew-zunger}
Electronic wave functions were represented using a plane wave basis set with a kinetic-energy cut-off of 35 Ry.
For 2D materials (monolayer, bilayer, and few-layer), we employed supercells with at least 16 \AA~of vacuum between slabs.
For the undistorted 1T structure, we sampled the Brillouin zone using grids of $32\times32\times16$ and $32\times32\times1$ k-points for bulk and few-layer systems, respectively.
For the C phase, $6\times6\times12$ and $6\times6\times1$ meshes were used for bulk and 2D systems, respectively.
An occupational smearing width of 0.002 Ry was used throughout.

Bulk structures were relaxed fully, including cell parameters as well as the atomic positions. For few-layer structures, the atomic positions and the in-plane lattice constants were optimized.
The frequencies and eigenvectors of zone-center phonon modes were calculated for the optimized structures using density functional perturbation theory, as implemented in Quantum Espresso.\cite{espresso}

\begin{figure}[]
\caption{
(Color online) An illustration of the Ta plane and Brillouin zone reconstruction under the lattice distortion of the C phase.
The blue cells represent the real space unit cell of the undistorted lattice (defined by the lattice vectors {\bf a} and {\bf b}) and the corresponding first Brillouin zone.
The green cells represent the super cell of the C CDW lattice (defined by the lattice vectors {\bf A} and {\bf B}) and the corresponding first Brillouin zone.
The dotted lines indicate the lateral shift of a supercell in an adjacent Ta plane under triclinic stacking.
}
\includegraphics[width=.85\linewidth]{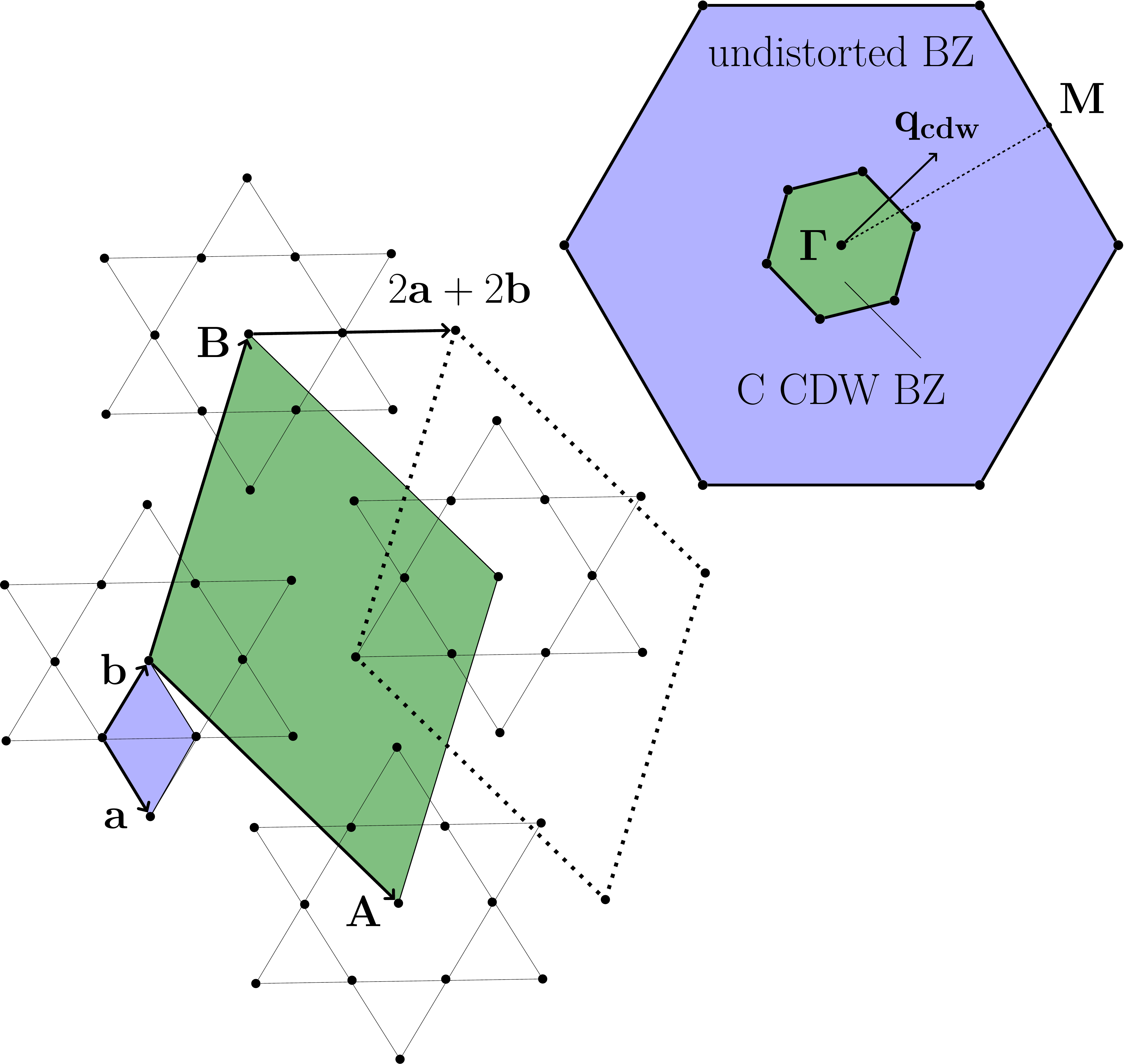}
  \label{fig:sod-triclinic}
\end{figure}

To investigate the effect of c-axis orbital texture in the bulk C phase, we considered two different supercells, referred to as the hexagonal and triclinic cells.
Both cells have in-plane lattice vectors $\mathbf{A} = 4\mathbf{a} + \mathbf{b}$ and $\mathbf{B} = -\mathbf{a} + 3\mathbf{b}$, where $\mathbf{a}$ and $\mathbf{b}$ are the in-plane lattice vectors of the undistorted cell, as shown in Fig. \ref{fig:sod-triclinic}.
The hexagonal cell has $\mathbf{C} = \mathbf{c}$, where $\mathbf{c}$ is perpendicular to the plane,  while the triclinic cell is described by $\mathbf{C} = 2\mathbf{a} + 2\mathbf{b} + \mathbf{c}$ (Fig. \ref{fig:sod-triclinic}).
In the triclinic cell, the central Ta atom in a cluster sits above a Ta in the outer ring of a cluster in the layer below.

\begin{figure}[b]
\caption{
(Color online) Optical modes of 1T bulk and monolayer structures: (a) E$_g$ and A$_{1g}$ modes (b)  E$_u$ and A$_{2u}$ modes.
Gray spheres represent Ta; yellow spheres represent S.
}
\includegraphics[width=.9\linewidth]{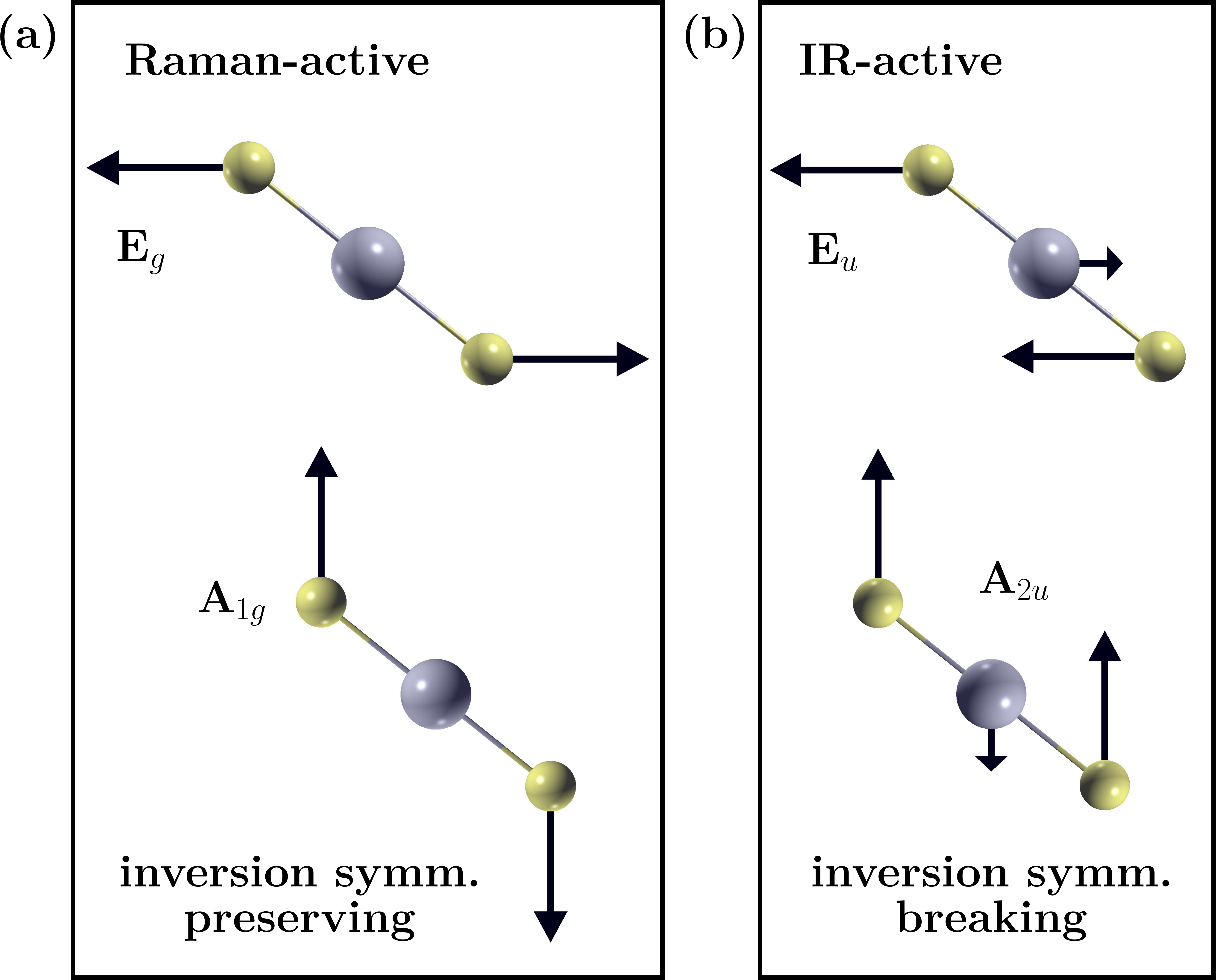}
\label{intracell}
\end{figure}

\subsection{\label{sec:exp-methods}Experimental Methods}

Multilayer 1T-TaS$_2$ was obtained by mechanical exfoliation onto Si/SiO$_2$ (300 nm) from bulk 1T-TaS$_2$ crystals synthesized via chemical vapor transport (CVT).\cite{liu2013superconductivity}
Optical microscopy was used to identify various thicknesses of exfoliated flakes, and atomic force microscopy (AFM) was used to measure flake thickness.
The Supplemental Material\cite{supplemental} contains the results and corresponding images for these measurements.
Raman spectra were collected via a Horiba LabRAM HR Evolution Raman spectrometer with 488 nm Ar/Kr wavelength laser and 1800/mm grating.
The laser power at the sample surface and the integration time were sufficiently low to prevent damage to the sample.
Back-scattered measurement geometry was adopted and the laser was focused on the sample through a 50$\times$ objective, with a 1 $\mu$m spot size.
The sample was kept in a sealed temperature stage, with pressure being less than 1$\times$10$^{-2}$ torr to minimize oxidation during the measurement.
The measurement temperature was controlled from 80 K to 600 K in a cold-hot cell with a temperature stability estimated to be $\pm 1$ K.
Polarized Raman spectra were collected with the relative angle of polarization between the incident and measured light ($\theta$) varied between 0 and 90$^\circ$.

\section{\label{sec:results&discussion}Results and Discussion}
\subsection{\label{sec:normal}Undistorted 1T-TaS$_2$}

The undistorted phase of bulk 1T-TaS$_2$ is stable above 550 K.
We did not measure Raman spectra in this temperature range because our samples showed signs of damage.
Nevertheless, a theoretical study of the vibrational modes is a useful starting point for understanding the more complicated C phase.

The point group of the undistorted 1T structure is $D_{3d}$ for both bulk and 2D crystals.
The zone-center normal modes belong to the representations $\Gamma = n(A_{1g} + E_g + 2A_{2u} + 2 E_u)$, where $n$ represents the number of layers except for the bulk, where $n=1$.
One of the A$_{2u}$ and one of the E$_u$ representations correspond to acoustic modes.
The two-fold degenerate E$_g$ and E$_u$ optical modes involve purely in-plane atomic displacements, while the A$_{1g}$ and A$_{2u}$ optical modes involve only out-of-plane displacements.
The displacement patterns of the optical modes for bulk and monolayer (1L) crystals are shown in Fig. \ref{intracell}.
Since the point group contains inversion, the A$_{1g}$ and E$_g$ modes, which preserve inversion symmetry, are Raman active, whereas the A$_{2u}$ and E$_u$ modes are IR active.

\begin{figure}[]
\caption{
  (Color online) Calculated phonon spectra of undistorted 1T-TaS$_2$: (a) $\Gamma$ phonons in few-layer crystals, (b) $\Gamma$ phonons in the bulk crystal, and (c) bulk phonons at $\mathbf{q_{cdw}}$ for hexagonal and triclinic stacking.
In (a) and (b), the doubly and singly degenerate modes  have E and A symmetry, respectively.
  In (c), the degeneracy corresponds to the number of equivalent $\mathbf{q_{cdw}}$ points in the Brillouin zone.
  Unstable modes (imaginary frequencies) at $\mathbf{q_{cdw}}$ are omitted.\\
}
\includegraphics[width=.9\linewidth]{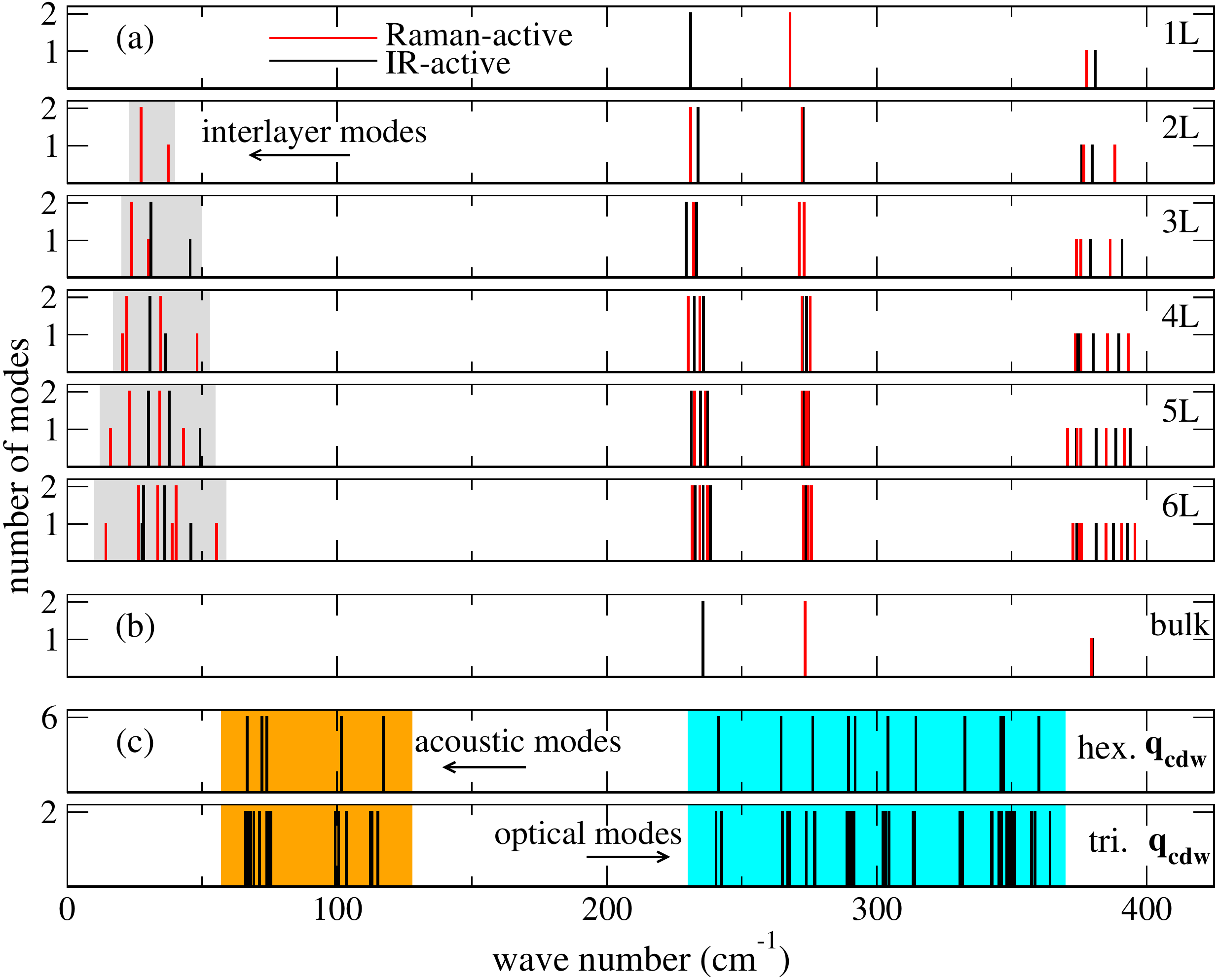}
\label{undistorted-spectrum}
\end{figure}

\begin{figure}[b]
\caption{
(Color online) Simplified illustration of the low-frequency phonon modes of a 1T bilayer: (a) interlayer shear and (b) interlayer breathing.
Gray spheres represent Ta; yellow spheres represent S.
The dots represent inversion centers.
}
\includegraphics[width=.9\linewidth]{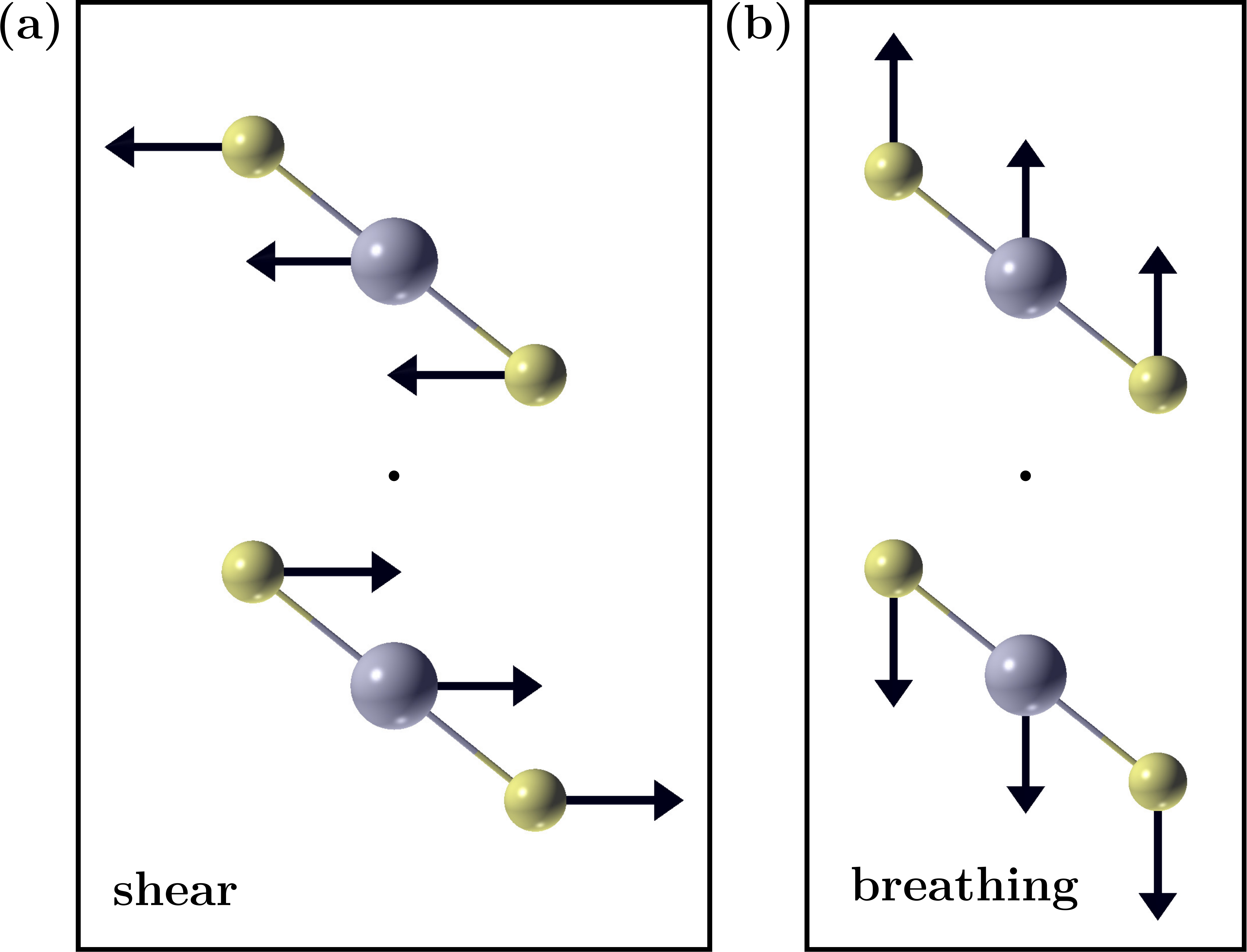}
\label{interlayer}
\end{figure}

The zone-center phonon frequencies calculated for monolayer and few-layer 1T-TaS$_2$ are plotted in Fig. \ref{undistorted-spectrum}(a), and those for the bulk are shown in Fig. \ref{undistorted-spectrum}(b).
The modes separate into three frequency ranges: acoustic and interlayer modes below about 50 cm$^{-1}$, in-plane E$_g$ and E$_u$ optical modes between about 230 and 270 cm$^{-1}$, and high-frequency out-of-plane A$_{1g}$ and A$_{2u}$ optical modes between about 370 and 400 cm$^{-1}$.
The low-frequency interlayer modes are present when $n>1$ and correspond to vibrations in which each layer displaces approximately rigidly, as illustrated in Fig. \ref{interlayer} for the bilayer (2L).

The 1T structure retains inversion symmetry regardless of the number of layers $n$.
This is in contrast to the 2H structure, in which the symmetry depends on whether $n$ is odd or even.
However, since the center of inversion for an $n$-layer 1T crystal alternates between being on a Ta site for odd $n$ and in-between TaS$_2$ layers for even $n$ (Fig. \ref{interlayer}), the displacement patterns of Raman-active modes differ depending on whether $n$ is even or odd.
We see this most clearly in the ultra-low-frequency regime.
The highest mode in this regime is always an interlayer breathing mode (Fig. \ref{interlayer}(b)) in which the direction of displacement alternates from layer to layer.
For odd $n$, this mode is IR active, while for even $n$ it is Raman active.
The same effect is evident in the high-frequency out-of-plane modes.

To facilitate interpretation of the calculated and measured Raman spectra of the C CDW materials in Sec. \ref{sec:cdw}, we show in Fig. \ref{undistorted-spectrum}(c) the phonon frequencies of bulk undistorted 1T-TaS$_2$ at the CDW wave vectors, $\mathbf{q_{cdw}}$, for hexagonal and triclinic stacking.\cite{gstuff}
Modes from the $\mathbf{q_{cdw}}$ and equivalent points fold back to the $\Gamma$ point in the Brillouin zone of the C supercell (Fig. \ref{fig:sod-triclinic}).\cite{lake}
For both types of stacking, we find unstable modes (not shown), indicating that the structure is dynamically unstable against a periodic distortion characterized by $\mathbf{q_{cdw}}$.
Based on Fig. \ref{undistorted-spectrum}(c), we expect that upon C CDW distortion, the folded-back acoustic modes will form a group near 100 cm$^{-1}$ (shown in orange), separated from the folded-back optical modes (shown in blue) by a gap of $\sim100$ cm$^{-1}$.
These are in addition to the original zone center modes.

\subsection{\label{sec:cdw}Commensurate CDW phase of 1T-TaS$_2$}

As in previous calculations,\cite{ritschel, millis, schwingenschlogl} we find the C phase energetically favorable compared with the undistorted structure for all thicknesses.
In the bulk, the system lowers in energy by $\sim9$ meV/TaS$_2$ and $\sim13$ meV/TaS$_2$ for hexagonal and triclinic stacking, respectively.
For a monolayer, the energy lowers $\sim23$ meV/TaS$_2$.

In the C CDW phase, the point group symmetry of a monolayer reduces from $D_{3d}$ to $C_{3i}$; inversion symmetry is preserved.
The normal modes at $\Gamma$ are classified as $19 E_g + 19 A_g + 20 E_u + 20 A_u$, with 57 Raman and 60 IR modes.
For bulk or few-layer C CDW with hexagonal stacking, the point group remains $C_{3i}$. With triclinic stacking (\emph{e.g.} $\mathbf{C} = 2\mathbf{a} + 2\mathbf{b} + \mathbf{c}$), the symmetry further reduces to $C_i$; only inversion symmetry remains, and the normal mode classification at $\Gamma$ becomes $n(57 A_g + 60 A_u)$.

\begin{figure}[]
\caption{
(Color online) Raman spectra of bulk ($>$ 300 nm) and thin (14.6 nm, 13.1 nm, 5.6 nm, and 0.6 nm) samples, measured at 250 K (NC phase) and 80 K (C phase).
All the spectra are normalized to their strongest Raman peak, indicated for the 0.6 nm curve by the arrow.
}
\includegraphics[width=\linewidth]{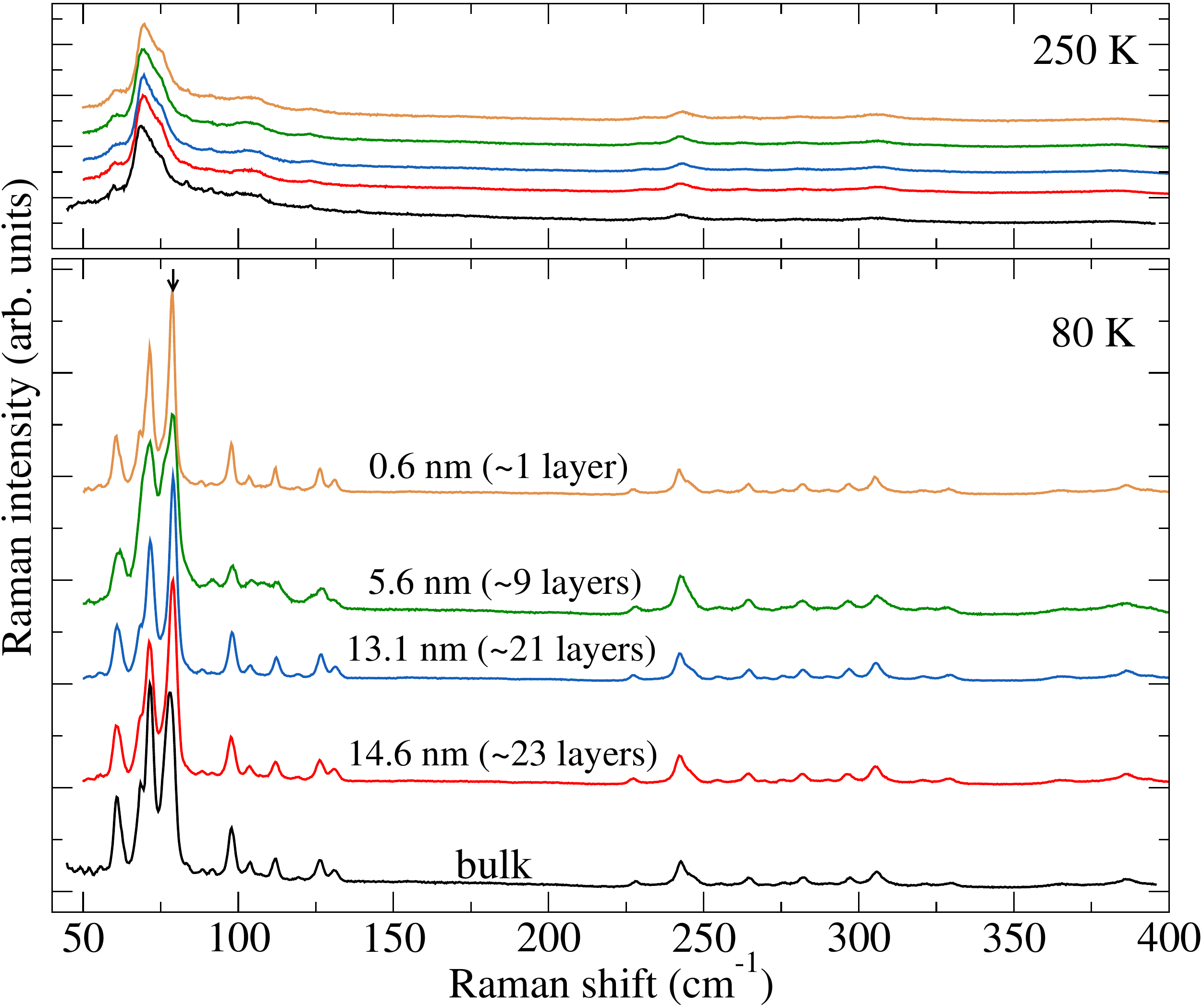}
\label{fig:raman-intensities}
\end{figure}

\begin{figure}[b]
\caption{
(Color online) (a) Raman lines measured for a bulk flake, at 80 K.
Calculated C phase Raman-active modes for bulk: (b) hexagonal stacking (c) triclinic stacking.
E$_g$ modes are doubly degenerate.
Triclinic stacking has A$_g$ modes only.
}
\includegraphics[width=\linewidth]{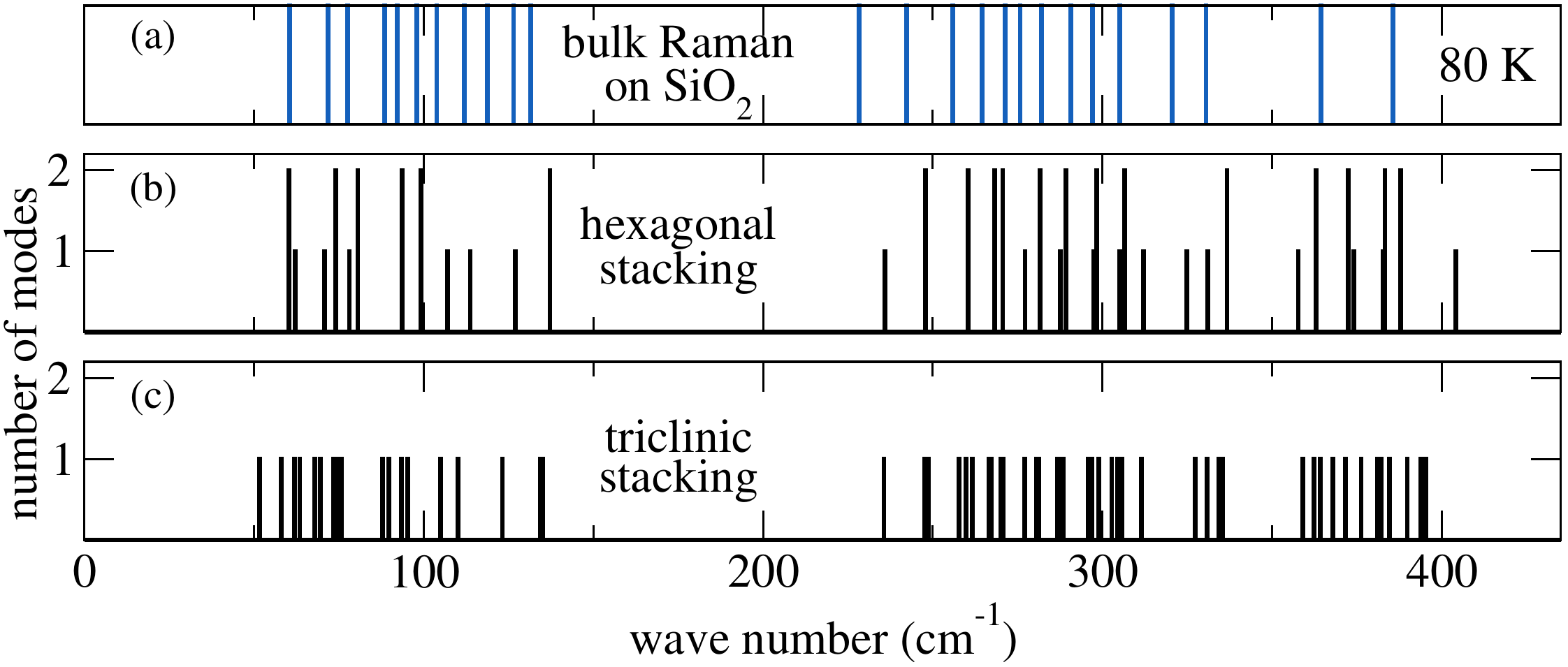}
\label{theory-experiment}
\end{figure}

The black curves in Fig. \ref{fig:raman-intensities} show the measured Raman spectrum of a bulk sample of 1T-TaS$_2$ on an SiO$_2$ substrate.
At 250 K (top panel), the sample is in the NC phase.
At 80 K (bottom panel), the sample is well below the NC--C transition temperature (180 K).
In going from NC to C, new peaks appear in the 60--130 cm$^{-1}$ and 230--400 cm$^{-1}$ ranges.
These spectra are consistent with previous Raman studies of the bulk material.\cite{duffey, hangyo, luican-mayer, sugai}

We calculated the phonon modes at $\Gamma$ for the bulk C structure with hexagonal and triclinic stacking.
Fig. \ref{theory-experiment} shows the Raman-active modes alongside the frequencies extracted from bulk Raman measurements of Fig. \ref{fig:raman-intensities}.
Our calculations and experiment are in general agreement in that both find two groups of Raman-active modes, one ranging from about 50-130 cm$^{-1}$ and the other from about 230-400 cm$^{-1}$.
These correspond to the folded-back acoustic and optic modes, as anticipated in Fig. \ref{undistorted-spectrum}.
(A more detailed comparison of the theoretical and experimental Raman lines is presented later in this section.)
In 1T-TaSe$_2$, unlike what we observe for 1T-TaS$_2$, the folded-back acoustic and optical groups overlap.\cite{lake}
The smaller mass of S compared to Se raises the frequencies of the optical group in TaS$_2$.
The presence of folded-back acoustic modes and thus the C phase is easier to identify in this material due to the frequency gap.

\begin{figure}[]
\caption{
(Color online) (a) Raman lines measured for a flake 0.6 nm thick, at 80 K.
Calculated C phase Raman-active modes: (b) monolayer, 3\% strain (c) monolayer, no strain (d) bilayer, hexagonal stacking (e) bulk, hexagonal stacking.
For clarity, red lines represent A$_g$ modes and black lines represent E$_g$ modes.
The Supplemental Material\cite{supplemental} contains the calculated C phase Raman-active modes for a bilayer with triclinic stacking.
}
\includegraphics[width=.9\linewidth]{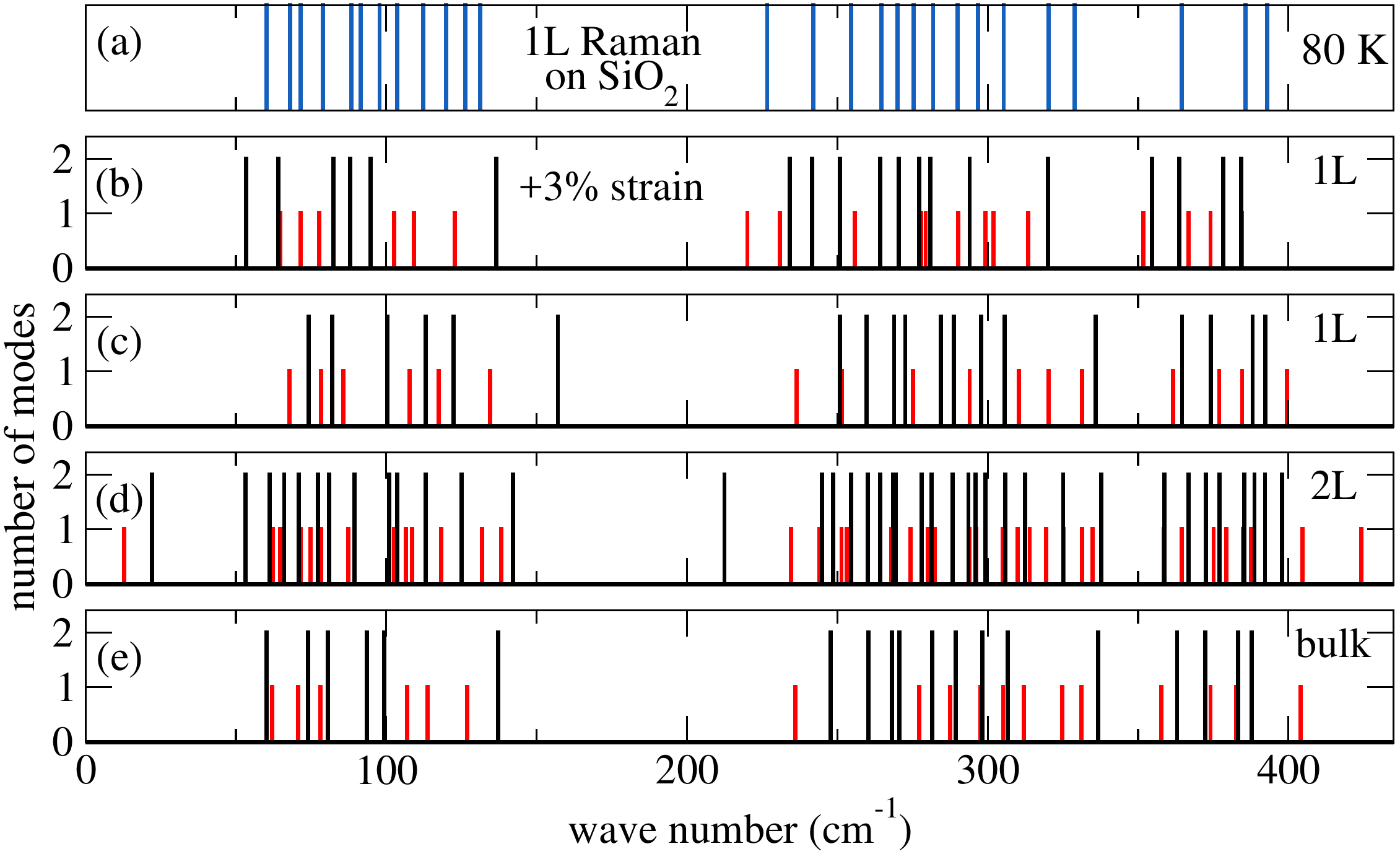}
\label{few-layer-distorted}
\end{figure}

\begin{figure}[]
\caption{
Measured polarized Raman spectra of the C phase for (top) a monolayer and (bottom) a 120 nm sample measured at 100 K.
We show the frequency range which gave the strongest signal.
The spectrum changes significantly between $\theta = 0^\circ$ (parallel polarization) and $\theta = 90^\circ$ (cross polarization).
Red and black arrows indicate peaks with and without polarization dependence, respectively.
The insets show comparisons of the measured polarized Raman lines and the calculated spectra.
}
\includegraphics[width=.9\linewidth]{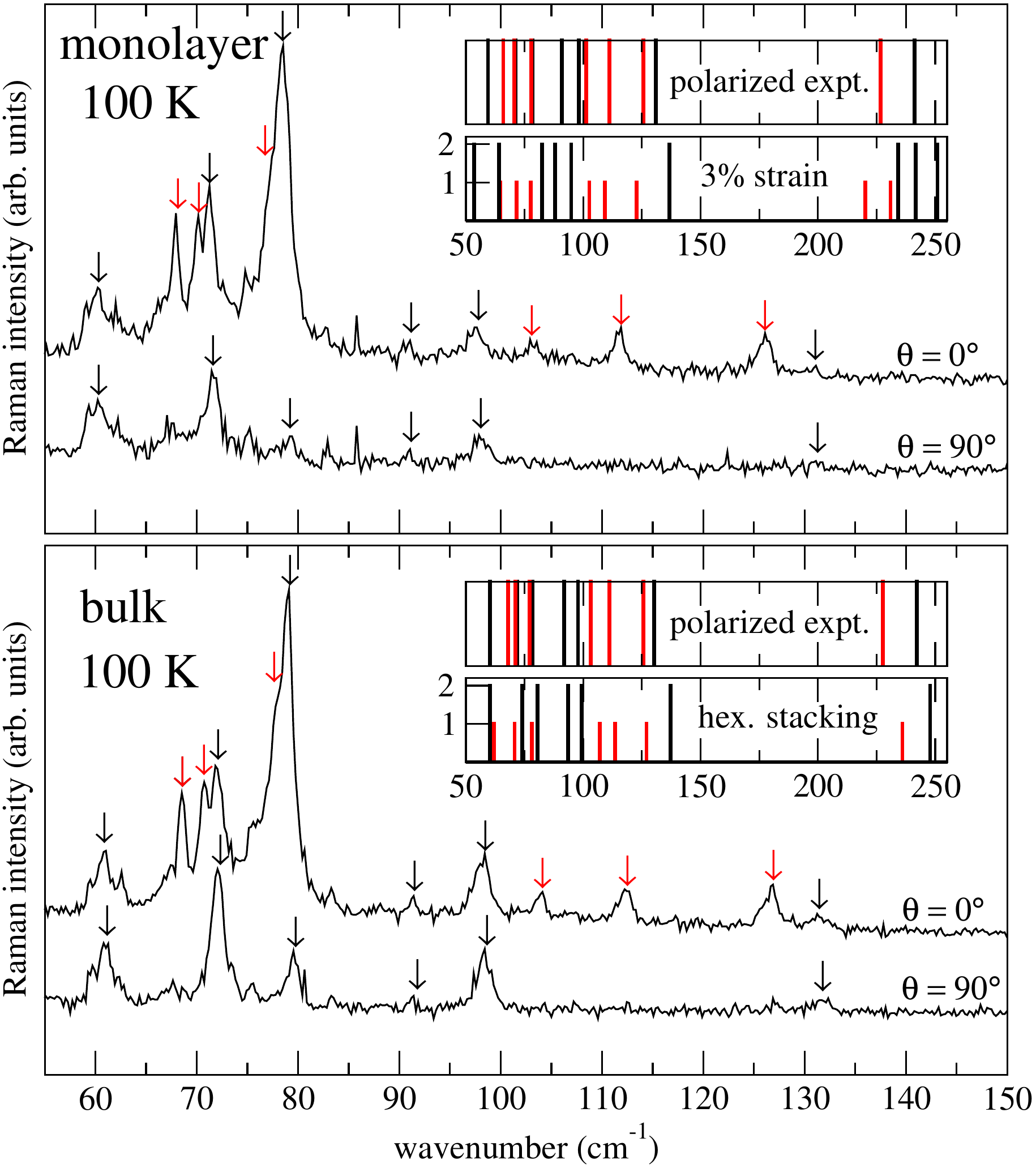}
\label{fig:polarized-raman}
\end{figure}

The red, blue, green and orange curves in Fig. \ref{fig:raman-intensities} are the measured Raman spectra for thin samples of 1T-TaS$_2$: 14.6 nm ($\sim$23 layers), 13.1 nm ($\sim$21 layers), 5.6 nm ($\sim$9 layers), 0.6 nm (1 layer).
As far as we are aware, this is the first report in the literature of the Raman spectrum of monolayer 1T-TaS$_2$.

We focus on the data taken at 80 K, where the bulk material is in the C CDW phase.
As the thickness is reduced from bulk to monolayer, the Raman spectrum remains largely unchanged, strongly suggesting that the C CDW structural phase persists as the samples are thinned, even down to a monolayer.
The monolayer spectrum is remarkably similar to that of the bulk (as well as to that of the 14.6 nm and 13.1 nm samples).
On the other hand, the spectrum of the 5.6 nm sample, while similar, shows some deviations.
In particular, some of the features are broadened, and there appears to be a higher background in the 80 - 130 cm$^{-1}$ range, which could be due to the presence of additional, closely spaced peaks.
These changes do not necessarily indicate a change in the C CDW lattice distortion.
In general, in going from a monolayer to a few-layer crystal, we expect weak interlayer interactions to shift and split each monolayer phonon line, as can be seen, for example, in our calculated results for the high-temperature undistorted phase of few-layer 1T-TaS$_2$ in Fig. \ref{undistorted-spectrum}.
The differences in the 5.6 nm ($\sim$9 layers) spectrum compared to those of the bulk-like and monolayer flakes (Fig. \ref{fig:raman-intensities}) could arise from such few-layer effects.

The thickness dependence of the NC to C phase transition has previously been investigated experimentally using transport,\cite{yu,tsen} electron diffraction,\cite{tsen} and Raman measurements.\cite{tsen2}
Most of these studies have reported difficulty in obtaining the C phase in thin samples, though, to our knowledge, none of them probed monolayer samples.
One previous Raman study suggested that the C phase is suppressed below thicknesses of about 10 nm at 93 K.\cite{luican-mayer}
It is not clear though, whether those samples were protected from surface oxidation, which can affect the transition.
Studies of thin 1T-TaS$_2$ samples that are environmentally protected by a h-BN overlayer indicate that the C phase persists down to 4 nm ($\sim$ 7 layers).\cite{tsen,tsen2}
In addition, the diffraction and transport measurements suggest the presence of large C phase domains in a 2 nm sample ($\sim$ 3 layers) at 100 K, although a conducting percolating path between the domains still exists.
The Raman spectra of this 2 nm sample at 10 K loses many of the well-defined peaks.\cite{tsen2}
Since our Raman measurements for the monolayer strongly indicate the C phase at 100 K, it may be that the barrier for the NC-C transition is smaller for the monolayer than for few-layer samples.
This would be consistent with the suggestion that there are distinct bulk and surface transitions, with the bulk having a larger activation barrier.\cite{tsen2}
Alternatively, it is possible that the encapsulating layer affects the transition in ultra thin samples.

The measured Raman spectra for the bulk and monolayer are remarkably similar, both in peak locations and relative peak heights.
Most of the discernible peaks shift by less than 1 cm$^{-1}$ in going from bulk to monolayer.
Our calculations, on the other hand, show a stronger dependence on dimensionality.
Fig. \ref{few-layer-distorted} shows the calculated 1L spectrum (c), alongside the hexagonally-stacked bulk spectrum (e).
To highlight mode shifts, we show the A$_g$ modes in red and the E$_g$ modes in black.
As in the experiments, the 1L and bulk spectra are similar, except the spectrum generally hardens upon thinning, particularly in the folded-back acoustic group, where the A$_g$ modes harden by $\sim$ 1 cm$^{-1}$ and the E$_g$ modes harden by up to $\sim$ 20 cm$^{-1}$.

This discrepancy between theory and experiment may be caused in part by approximations made in our DFT calculations.
For similar materials, LDA frequencies are typically within 5-10 cm$^{-1}$ of experiment.\cite{liang}
In addition, it is likely that substrate-induced effects, such as strain and screening, which are not taken into account in our calculations, are important.

In general, positive strain and enhanced screening tend to soften vibrational frequencies.
In the absence of a substrate, the reduced screening in a monolayer tends to harden the phonon modes, as seen in our calculations.
The presence of a dielectric substrate in the experiments mitigates this effect.
To explore the effect of strain, we calculated the Raman modes (Fig \ref{few-layer-distorted}(b)) for a 1L under 3\% strain as an example.
Straining the monolayer in our calculations generally softens the modes, bringing the folded-back acoustic group into better agreement with the measurement, while over-softening the folded-back optical group.
Since we did not include substrate screening in our simulations, the relative importance of strain and screening remains unclear.

As with the undistorted case in Sec. \ref{sec:normal}, the bilayer (Fig. \ref{few-layer-distorted}(d)) has ultra-low frequency modes, associated with interlayer shear and breathing.
The breathing mode (A$_g$) is calculated to be $\sim 30$ cm$^{-1}$ lower than in the undistorted case, indicating that the coupling between layers is weaker in the C phase; indeed, the structure swells by 0.1 \AA~in the distortion.
In comparing Fig. \ref{few-layer-distorted}(c) and (d), two 2L Raman modes stand out above and below the 1L optical group range: an E$_g$ mode around 210 cm$^{-1}$ and a A$_g$ mode around 425 cm$^{-1}$.
These modes evolve from IR-active 1L modes near 215 cm$^{-1}$ and 415 cm$^{-1}$ respectively (not shown).
As discussed in Sec. \ref{sec:normal} for the undistorted case, the modes can switch from IR to Raman depending on whether $n$ is even or odd.

Polarized Raman measurements differentiate modes of different symmetries.
For the low-temperature phase of 1T-TaS$_2$, this allows for a more detailed comparison of theory and experiment while potentially distinguishing between different c-axis orbital textures.
In the C phase monolayer and hexagonally-stacked bulk, the A$_g$ and E$_g$ modes depend differently on the relative angle $\theta$ between the polarization of the incident and scattered light.
In back-scattering geometry with parallel polarization ($\theta=0$), both A$_g$ and E$_g$ modes are detectable.
If incident and scattered light are cross polarized ($\theta=90^\circ$), the intensity of the A$_g$ modes become zero, leaving only the E$_g$ modes.
In contrast, with triclinic stacking of the bulk C phase, all Raman modes have the same symmetry, meaning they should all have the same dependence on $\theta$.

In Fig. \ref{fig:polarized-raman}, we plot the measured polarized Raman spectra for bulk and monolayer samples at 100 K comparing parallel and cross-polarized configurations.
These plots focus on the frequency range of the folded-back acoustic modes since these modes provide the clearest signature of the C phase.
The results are similar for the bulk and monolayer, and agree well with an early polarized Raman study.\cite{sugai}

For both the bulk and monolayer, we measure two types of modes with different $\theta$ dependence, consistent with A$_g$ and E$_g$ modes.
A line by line comparison of the measured bulk frequencies shows strong agreement with the A$_g$ and E$_g$ modes calculated for hexagonal stacking.
Most modes are a few cm$^{-1}$ higher than the measured frequencies, as is typical within LDA, which tends to overbind the lattice.\cite{liang}
For the monolayer, neither the strained nor unstrained calculations match the measured spectrum as closely.
We conclude that the primary reason is a lack of substrate effects in our calculations.

The good agreement with the calculated A$_g$ and E$_g$ modes for hexagonally-stacked bulk is surprising since diffraction data do not support pure hexagonal stacking in bulk samples.\cite{wilson, williams, scruby}
It is likely that the c-axis orbital texture is more complex, involving both hexagonal and triclinic alignment between adjacent layers, as suggested by Ref. \citenum{nakanishi}.
Our data are inconsistent with a random or triclinic stacking unless the individual layers are so weakly coupled in the bulk that they vibrate like nearly independent monolayers.\cite{sugai}

\section{\label{sec:discussion}Conclusions}

We have calculated the phonon modes of both the undistorted and C phases of 1T-TaS$_2$ in the bulk and ultrathin limits.
We also measured the C phase Raman spectrum in bulk and thin samples.
Through our calculations, we identified the low-frequency folded-back acoustic Raman modes as a signature of the C phase; these modes conveniently lie in a low-frequency range that does not overlap with optical modes, and in which there are no zone-center modes in the undistorted phase and very few in the NC phase.
In our measured Raman spectra, these modes persist down to 1L, which is below the critical thickness previously suggested for the NC-C transition.\cite{yu}
The NC phase consists of C phase domains separated by metallic interdomain regions.
As the temperature is lowered, the domains grow as the transition is approached.
Tsen, \emph{et al.} observed commensuration domains $\sim500$ \AA ~in length at 100 K in a 2 nm-thick sample ($\sim$ 3 layers).\cite{tsen}
Our measurements show that at 100K, the monolayer has transformed into the C phase, or at least has C domains larger than our laser spot size of 1 $\mu$m.
This suggests that the barrier for the NC-C transition is lower in the monolayer than in few-layer samples, though further investigation is needed, particularly regarding the impact of substrates and overlayers on the transition.

Recently, there has been interest in the effect of c-axis orbital texture on conduction, with the hope of controlling it.\cite{ritschel}
It would be useful to have a way to quickly identify the c-axis order in these materials.
Since different stackings have different symmetries, comparison of cross-polarized and parallel-polarized Raman experiments could provide information about the orbital texture of the bulk or few-layer crystals.
For example, Ritschel, \emph{et al.} have proposed a 2L 1T-TaS$_2$ device operating in the C phase.
The `on' state would be conducting with triclinic stacking, the `off' state would be insulating with hexagonal stacking, and the switching mechanism could perhaps be an external field.\cite{ritschel}
In testing the proposed switching mechanism, polarized Raman could provide an easy way to determine a change in the orbital texture of the sample.

More generally, our analysis of the 1T vibrational modes complements the existing literature that focuses on the 2H phase of TMDs.
Although the 2H structure is more common, synthesis of metastable 1T phases has been investigated.\cite{ma2015reversible, kappera2014phase, voiry2015covalent}
Besides aiding interpretation of the Raman modes of C phase 1T-TaS$_2$, our analysis of the undistorted phase is more broadly applicable to other 1T materials.

\begin{acknowledgments}
This work was supported by NSF Grants DMR-1358978 \& EFRI-143307.
The authors acknowledge TACC and SDSC for HPC resources provided through an XSEDE allocation.
J. Freericks was also supported by the McDevitt Bequest at Georgetown University.
J. Robinson would like to acknowledge Prof. Y. P. Sun for providing bulk 1T-TaS$_2$ crystals.
\end{acknowledgments}


\begin{thebibliography}{52}
\bibitem [{gst()}]{gstuff}%
  \BibitemOpen
  \href@noop {} {}\bibinfo {note} {For the hexagonal stacking, two inequivalent
  reciprocal lattice vectors of the CDW cell lie within the Brillouin zone of
  the undistorted cell: $\mathbf{g_1}$ and $\mathbf{g_1} + \mathbf{g_2}$. Each
  is 6-fold degenerate. For triclinic stacking, there are 6 such vectors:
  $\mathbf{g_1}, \mathbf{g_2}, \mathbf{g_1} - \mathbf{g_2}, 2\mathbf{g_1} -
  \mathbf{g_2}, \mathbf{g_1} - 2\mathbf{g_2}$, and $\mathbf{g_1} + \mathbf{g_2}
  + \mathbf{g_3}$. Each is 2-fold degenerate.}\BibitemShut {Stop}%
\end{thebibliography}
\end{document}